Taylor & Francis
Taylor & Francis Group

# What is the market potential for on-demand services as a train station access mode?

Nejc Geržinič , Oded Cats , Niels van Oort , Sascha Hoogendoorn-Lanser  and Serge Hoogendoorn

Department of Transport and Planning, Delft University of Technology, Delft, Netherlands

**ABSTRACT**

On-demand mobility services (FLEX) are often proposed as a solution for the first/last mile problem. We study the potential of using FLEX to improve train station access by means of a three-step sequential stated preference survey. We compare FLEX with the bicycle, car and public transport for accessing two alternative train stations. We estimate a joint access mode and train station choice model. Estimating a latent class choice model with different nesting structures, we uncover four distinct segments in the population. Two segments (∼ 50%) with a lower Willingness-to-Pay seem to be more likely to take-up FLEX. Ex-urban car drivers seem to be the most likely segment to adopt FLEX, showing great, since members of this segment are currently frequent users of the private car. Our case study also shows that while FLEX competes primarily with public transport when accessing local stations, it competes primarily with car for reaching distant stations.

## 1. Introduction

Train travel is acutely affected by the first/last mile problem. With a significant number of people not living within walking distance of a train station, it is clear that train travel is heavily dependent on how well travellers can access and egress the train station (Brons, Givoni, and Rietveld 2009). The most common train station access mode (on the home-end) in the Netherlands is the bicycle, representing almost half of all trips (Shelat, Huisman, and van Oort 2018), with walking and local public transport (bus, tram and metro or BTM) accounting for around 15% each and the rest being either as a car driver, car passenger or other modes. Similar to walking, cycling is strongly impeded by distance, with the attractiveness decreasing significantly for distances above three kilometres (Keijer and Rietveld 2000), at which point motorised modes like public transport and car become comparatively more attractive.

In recent years, on-demand services (both flexible public transport and ride-hailing services like Uber or Lyft) have begun operating, also as first/last-mile access to mass transit (Phun, Kato, and Chalermpong 2019), with the advent of smartphone technology further

**CONTACT** Nejc Geržinič ✉ n.gerzinic@tudelft.nl 🏢 Department of Transport and Planning, Faculty Of Civil Engineering and Geosciences (Stevinweg 1, 2628 CN Delft), Delft University of Technology, Delft, Netherlands boosting the rapid emergence and deployment of such services. Several studies assert that on-demand services both attract passengers from public transport services and at the same time act as an access/egress providers to public transport stations (Alemi et al. 2018; Clewlow and Mishra 2017; Deka and Fei 2019; Hall, Palsson, and Price 2018; Sikder 2019; Tirachini 2019; Tirachini and del Río 2019; Young, Allen, and Farber 2020; Young and Farber 2019).

Research shows that the longer their trip, the longer travellers are willing to travel on their access mode. According to Krygsman, Dijst, and Arentze (2004), access and egress time can account for up to 50% of total travel time. While for longer trips the overall access and egress times are longer, they account for a lower share of the total trip time. Travel time spent travelling with access modes is predominantly found to be valued higher (perceived more negatively) than travel time on the main leg of the trip (Arentze and Molin 2013; Bovy and Hoogendoorn-Lanser 2005; La Paix Puello and Geurs 2014). Travel time on the access leg is also found to be a key determinant, both for station access mode choice (Halldórs-dóttir, Nielsen, and Prato 2017; van der Waerden and van der Waerden 2018) as well as for airport ground access (Jou, Henshel, and Hsu 2011), where both in-vehicle and out-of-vehicle time components were found to be crucial in mode choice. In order to increase the catchment area of train stations beyond the current range of active modes, improving the quality of (public/shared) motorised access modes is therefore essential.

Past findings based on transit ridership data (Hall, Palsson, and Price 2018), household travel behaviour surveys (Clewlow and Mishra 2017) and intercept surveys (Rayle et al. 2016; Tirachini and del Río 2019) suggest that on-demand services generally reduce the ridership of local bus and light rail transport and increase the ridership of longer-distance rail services. Tirachini and del Río (2019) find that for every user that accesses public transport with a ridesourcing service, eleven users switch from public transport. The authors argue that this is not necessarily entirely negative, as in the latter case, travellers are infrequent public transport users and the trips happen at the edges of the day, when public transport services are often limited. In contrast, findings by Dong and Ryerson (2020) on airport ground access suggest that the entry of Uber and Lyft onto the market has primarily impacted taxi trips, with transit ridership seeing a very limited impact based on the trend.

Ridesharing and ridesourcing services have the potential to provide first/last mile connectivity to public transportation. The potential of the former is explored by Stiglic et al. (2018), who analysed peer-to-peer ridesharing (different from ride-hailing from an organisational perspective, but very similar for the passenger) where drivers (themselves commuters) would pick up passengers along the way and drop them off at a train station, potentially also parking there and taking the train themselves. They report an improvement in the matching rate both when ridesharing is offered as station access instead of only for the entire trip, as well as by allowing the driver to pick up two passengers, instead of just one. On-demand services could be subsidised to make them more affordable, increase their attractiveness and thereby also the attractiveness of public transport. Reck and Axhausen (2020) find that the travel time saved by using ridesourcing rather than walking does not outweigh the additional cost and transfer. This could be due to the rather short access distances in the data (with an average of 1–1.5 km). The authors suggest that over longer access distances and especially if a transfer can be saved on the public transport leg, using ridesourcing as an access mode could prove beneficial. Taxi (on-demand) services were also found to be attractive for a majority of people accessing high speed railway stations in Taiwan (Wen, Wang, and Fu 2012).

Access mode choice is often only one part of a larger choice process, as passengers may be located in the vicinity of more than one train station and therefore also have to choose which station to access for their trip. The attractiveness of stations is determined on one hand by their facilities (e.g. parking availability, shops, ticket counters) and on the other hand by the rail service quality. The latter was defined by Debrezion, Pels, and Rietveld (2009) as the Rail Service Quality Index (RSQI), which is based on the (1) frequency of the service / waiting time at the station, (2) connectivity of that station in the network (number of transfers needed to destinations), (3) location in the network (travel time to destinations) and (4) the price to reach those destinations. They then used this RSQI to estimate a combined access mode and station choice based on revealed preference (RP) data from the Netherlands. With respect to station characteristics, they conclude that indeed both rail services and (parking) facilities at stations significantly increase the station's attractiveness. For access mode choice, their findings are in line with the literature in that cycling and especially walking are highly affected by the access distance, with public transport being least sensitive to the distance. Joint mode and station choice was also researched by Bovy and Hoogendoorn-Lanser (2005), who characterised the train services based on the travel time, number of transfers and the type of service as either InterCity (IC) or local trains only. While the former two attributes were determined to be significant, the latter was not. The authors speculate that this is a consequence of their focus on shorter trips. Comparing the travel time estimates, in-vehicle time (IVT) on the train was found to be perceived less negatively than access time by private modes (bike and car), but more negatively than public transport access time. The respective weights for the two access IVT components were reported as 1.6 and 0.8 compared to the train IVT. Transfers were also found to have a significant impact, with higher frequency ($> 6x/h$) transfers having a lower impact than low frequency ($\leq 6x/h$) transfers. Travel time, service frequency and parking availability were also found to be significant predictors of station choice by Chakour and Eluru (2014) and by Fan, Miller, and Badoe (1993). Chakour and Eluru (2014) concluded that improvements in access time (especially for public transport and active modes) largely impacts mode choice and not station choice. Fan, Miller, and Badoe (1993) modelled car and public transport access separately, reporting that travellers who travel by car, perceive travel time less negatively and attach greater value to the frequency of train services compared to travellers who access train stations by public transport.

When modelling the joint access-mode-and-train-station choice, a nesting structure is often included in the model specification. This enables the model to capture correlations between (unobserved) utilities of alternatives which are modelled in the same nest. With the estimation of access and station choice, two possible nesting structures can be formed, where either the station is chosen first (station-based nesting) or the access mode is chosen first (mode-based nesting). Studies report mixed outcomes, with some finding that station-first models achieve a better model fit (Bovy and Hoogendoorn-Lanser 2005; Chakour and Eluru 2014), whereas others concluding that mode-first models prove superior (Debrezion, Pels, and Rietveld 2009; Fan, Miller, and Badoe 1993). Interestingly, in a study of joint access mode and airport choice in the New York City area (Gupta, Vovsha, and Donnelly 2008), a model without any nesting structure was found to be superior. While these results are also influenced by the exact context of the SP and RP data, most studies find the differences between the models to be relatively small.

The behavioural characteristics of passengers' choices in the context of accessing larger transportation hubs, i.e. a train stations, airports etc. has been widely studied. More recently, advancements have also been made in understanding how on-demand services impact travel behaviour in both urban and rural areas, due to their on-demand nature, potential pooling with other passengers, detours and time variability etc. Notwithstanding, to the best of our knowledge, the intersection of these two topics, i.e. the behavioural preferences of accessing public transportation by means of on-demand mobility, remains unknown, despite their growing relevance in the urban mobility landscape worldwide. Although this topic has been somewhat studied for airport ground access, Gupta, Vovsha, and Donnelly (2008) state that airport access trips are highly specific and differ significantly to typical commute trips, making the generalisation of their results difficult.

Our study fills the aforementioned research gap, providing insight into how on-demand services can be utilised as an access mode for train stations, as well as how this may impact station choice of travellers. We carry out a stated preferences survey of joint access mode and train station choice. The contributions of this study are threefold: (1) highlighting the preferences of travellers associated with on-demand mobility services, (2) estimating how the characteristics of the access leg and the train leg are traded off and (3) segmenting the population based on their preferences towards on-demand mobility, train station access and the nesting structure that best captures the joint mode-station choice.

The rest of the paper is structured as follows: the survey design, model estimation and data collection are described in the Section 2. The results of the analysis and the uncovered latent segments are then presented in Section 3. Section 4 demonstrates four different scenarios of introducing on-demand services and how those could impact the modal split, and presents the sensitivity of users to certain design aspects. The findings are then summarised and their policy implications discussed in Section 5.

## 2. Methodology

To analyse the potential impact of on-demand services on passenger train station choice, a stated preference survey is carried out in which both access mode choice and station choice are evaluated. The design of the survey is outlined in Section 2.1. Several choice models are then estimated, to gain an understanding of the respondents' travel behaviour preferences, as described in Section 4. Finally, the data collection is presented in Section 2.3.

### 2.1. Survey design

Although several smaller scale on-demand services are operating in the Netherlands (Bronsvoort et al. 2021), most people are not yet familiar with this type of service. Thus, a stated choice experiment is chosen to obtain travel preference information. To capture both the access mode and train station choice, a three-step sequential stated preference survey is carried out (Choudhury et al. 2018), as shown in Figure 1. In the first two steps (Choice 1 and Choice 2), respondents choose one of five available modes to access stations A and B (four modes if they do not have a driving licence and access to a car). The third choice then integrates information on the access modes for each station as chosen by the respondents and the train service characteristics of that particular station. This choice process is repeated for a total of six hypothetical trips. Examples of the choice sets for all three choices are also shown in Appendix A.

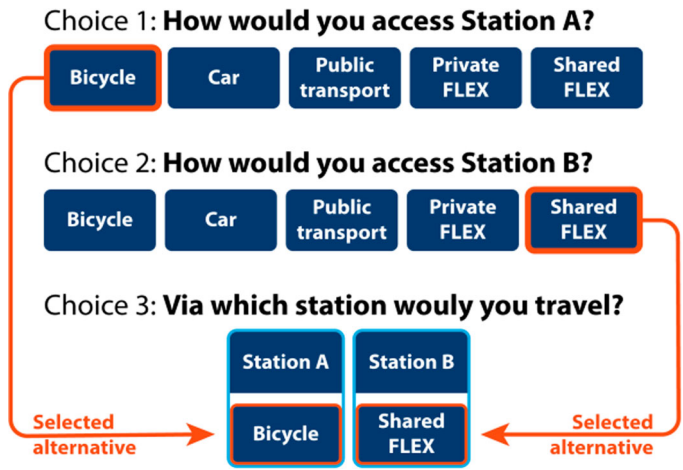

**Figure 1.** Survey outline for the three choices within one choice set.

Train station access/egress trips can be on the home-end or activity-end of the train journey. Activity-end trips are interesting for the potential of shared modes as travellers commonly do not have a private mode available and thus must use a shared mode of transport. The most common activity-end modes in the Netherlands are walking and public transport (MRDH 2016; Stam et al. 2021), with shared/micro- mobility becoming a more and more attractive alternative. Home-end trips can be interesting from the perspective of on-demand mobility, as these trips tend to be longer than activity-end trips (MRDH 2016; Stam et al. 2021), meaning people are less likely to walk and therefore more likely to choose a bicycle or a motorised mode of transport. While the availability of private modes on the home-end reduces the attractiveness of on-demand mobility, it can become more attractive when coupled with policies that restrict car use, such as higher parking cost and fewer car parking spots. Although both sides of a train trip are interesting from an on-demand mobility perspective, this study looks into the home-end of the train trip.

To that end, the survey includes three of the most frequently used access modes at the moment: bicycle, car, public transport (specified as either bus, tram or metro) (MRDH 2016; Stam et al. 2021) and two on-demand service options, a private and a shared service. The on-demand service is branded as FLEX, to ease communication and because this name is often used in the Netherlands for such services (Bronsvoort et al. 2021). Each of the access modes is characterised by three attributes: (1) cost, (2) (in/on-vehicle) travel time and (3) out-of-vehicle (OVT) time. Cost refers to the trip cost (car, PT, FLEX) and parking cost (bicycle, car). In-vehicle time is the time spent on the move and only includes time in (on-board) the vehicle. Out-of-vehicle time is defined as (a) 'parking search time and time walking to the station' for bicycle and car, as (b) 'walking to a nearby PT stop and waiting' for public transport and as (c) 'waiting (at home)' for the two FLEX alternatives. The attributes and their corresponding levels are presented in Table 1.

**Table 1.** SP survey prior values and attribute levels.

| Access leg prior | | | Train leg priors | | |
|---|---|---|---|---|---|
| Cost | | −0.6 | Cost | | −0.6 |
| In-vehicle time | | −0.1 | In-vehicle time | | −0.1 |
| Out-of-vehicle time | | −0.2 | Headway | | −0.1 |
| | | | Transfer | | −1.2 |
| | | | InterCity station label | | 0.7 |

| | Cost | | In-vehicle time | | Out-of-vehicle time | |
|---|---|---|---|---|---|---|
| | Local station | InterCity station | Local station | InterCity station | Local station | InterCity station |
| Bicycle | € 0.00 | € 0.00 | 12 min | 30 min | 1 min | 1 min |
| | € 1.00 | € 1.00 | 16 min | 35 min | 5 min | 5 min |
| | € 2.00 | € 2.00 | 20 min | 40 min | 9 min | 9 min |
| Car | € 1.00 | € 2.00 | 8 min | 12 min | 1 min | 1 min |
| | € 5.00 | € 6.00 | 12 min | 21 min | 5 min | 5 min |
| | € 9.00 | € 10.00 | 16 min | 30 min | 9 min | 9 min |
| Public transport | € 0.50 | € 1.00 | 8 min | 12 min | 1 min | 1 min |
| | € 2.00 | € 3.00 | 12 min | 21 min | 5 min | 5 min |
| | € 3.50 | € 5.00 | 16 min | 30 min | 9 min | 9 min |
| Private FLEX | € 5.00 | € 8.00 | 8 min | 12 min | 1 min | 1 min |
| | € 10.00 | € 13.00 | 12 min | 21 min | 5 min | 5 min |
| | € 15.00 | € 18.00 | 16 min | 30 min | 9 min | 9 min |
| Shared FLEX | € 2.00 | € 2.00 | 8 min | 12 min | 1 min | 1 min |
| | € 5.00 | € 6.00 | 12 min | 21 min | 5 min | 5 min |
| | € 8.00 | € 10.00 | 16 min | 30 min | 9 min | 9 min |

| | Cost | In-vehicle time | Headway | Transfers |
|---|---|---|---|---|
| Local station | € 17.00 | 60 min | 10 min | 1 |
| | € 20.00 | 75 min | 15 min | 2 |
| | € 23.00 | 90 min | 30 min | 3 |
| InterCity station | € 20.00 | 75 min | 10 min | 0 |
| | | | 15 min | 1 |
| | | | 30 min | |

The station choice is the third and final step of the choice process, where the respondents are shown their selected access modes and attributes, along with four characteristics of the train services at the respective station. The access distance used to determine the attribute levels for the two stations are approximately five and ten kilometres from the trip origin (home). As five kilometres is the average access distance to the nearest train station in the Netherlands (CBS 2011), this is an appropriate access distance to assume. The inclusion of a second, more distant station, is made with the goal of testing if respondents are willing to travel further for a different train service and if different access modes are offered. According to the Dutch Statistics Bureau (CBS 2011), the average distance to the nearest interchange station (offering potentially a more direct service) in the Netherlands is 10.5 km. To avoid respondents having an inherent preference for either of the stations, they are only labelled as 'Station A' and 'Station B' in the experiment. Given the access distances, we refer to them from here on as the 'Local station' and 'Distant station' respectively. Based on results from literature (Debrezion, Pels, and Rietveld 2009; van Mil et al. 2021), we characterise the train stations and services by (1) the trip cost (only for the train leg), (2) total travel time on the train(s), including the transfers, (3) train service headway and (4) the number of transfers on the train leg of the trip. The attributes and levels are summarised in Table 1.

A D-efficient design with six choice sets is constructed in Ngene (ChoiceMetrics 2018), with prior parameter values obtained from the literature. The prior values (found in Table 1) are determined based on the value of travel time of 10 €/h in the Netherlands (Kouwenhoven et al. 2014). From that, we specify the IVT prior as −0.1 and the cost prior as −0.6. Priors for other attributes are based on IVT-equivalent minutes (multipliers) reported in the literature (Arentze and Molin 2013; Bovy and Hoogendoorn-Lanser 2005; Frei, Hyland, and Mahmassani 2017; Wardman 2001, 2004). With respect to mode specific constants, we found a large range of preferences (Arentze and Molin 2013; Bovy and Hoogendoorn-Lanser 2005; Choudhury et al. 2018; Currie 2005; Frei, Hyland, and Mahmassani 2017; Paleti et al. 2014; Rose and Hensher 2014), differing not only in their relative preference (compared to IVT), but also in the order of which modes are preferred over others. Hence, we decide not to specify any prior values for the Alternative Specific Constants (ASCs).

To get insights into the attitudes towards new mobility services, respondents are asked to respond to 16 Likert-type questions (shown in Table 2). The statements are associated with different characteristics of FLEX services, based on the categories defined by Durand et al. (2018): (1) Use of smartphone apps, (2) Mobility integration, (3) Sharing a ride and (4) Sharing economy. They are also asked to indicate their familiarity with six service of the sharing economy, four of which are in the mobility domain (found in Table 3). Additional socio-demographic and travel behaviour information is obtained from other surveys in the Dutch Mobility Panel (Hoogendoorn-Lanser, Schaap, and Oldekalter 2015).

## 2.2. Model estimation

We analyse the obtained SP observations by estimating a series of choice models using the PandasBiogeme package for Python (Bierlaire 2020). The data is analysed under the assumption that respondents make decisions by maximising their perceived utility (McFadden 1974). Given the nature of the 3-step stated choice experiment, different model specifications were tested for how to include the alternatives of the different choice steps in the model estimation. In the end, the most realistic option, capturing the characteristics of all available alternatives, is used. The model is made up of 10 alternatives, consisting of 5 alternatives for each of the 2 train stations (8 alternatives in total for respondents without a driver's licence or an access to car).

**Table 2.** Attitudinal statements on FLEX-related characteristics.

| Category | | Statement |
|---|---|---|
| Use of (travel planning) apps | 1 | I find it difficult to use travel planning apps.[a] |
| | 2 | Using travel planning apps makes my travel more efficient.[a] |
| | 3 | I am willing to pay for transport related services within apps. |
| | 4 | I do not like using GPS services in apps because I am concerned for my privacy. |
| Mobility integration | 5 | I am confident when travelling with multiple modes and multiple transfers. |
| | 6 | I do not mind infrequent public transport, if it is reliable. |
| | 7 | I do not mind having a longer travel time if I can use my travel time productively. [b] |
| | 8 | Not having to drive allows me to do other things in my travel time. [b] |
| Sharing a ride | 9 | I am willing to share a ride with strangers ONLY if I can pay a lower price. [b] |
| | 10 | I feel uncomfortable sitting close to strangers. [b] |
| | 11 | I see reserving a ride as negative, because I cannot travel spontaneously. |
| Sharing economy | 12 | I believe the sharing economy is beneficial for me. |
| | 13 | I believe the sharing economy is beneficial for society. |
| | 14 | Because of the sharing economy, I use traditional alternatives (taxis, public transport, hotels …) less often. |
| | 15 | Because of the sharing economy, I think more carefully when buying items that can be rented through online platforms. |
| | 16 | I think the sharing economy involves controversial business practices (AirBnB renting, Uber drivers' rights …). |

[a]Adapted from (Lu et al. 2015).
[b]Adapted from (Lavieri and Bhat 2019).
The remaining statements were formulated for the purpose of this study.

**Table 3.** Service of the sharing economy, including examples, as presented to respondents.

| | Type of (sharing economy) service | Examples shown |
|---|---|---|
| 1 | How familiar are you with car sharing? | Snappcar, Greenwheels, car2go |
| 2 | How familiar are you with bike/scooter sharing? | Mobike, OV fiets, Felyx |
| 3 | How familiar are you with flexible public transport? | Twentsflex, Bravoflex, U-flex, Delfthopper |
| 4 | How familiar are you with ride-hailing? | Uber, ViaVan |
| 5 | How familiar are you with food delivery services? | Thuisbezorgd, Deliveroo, Foodora, UberEATS |
| 6 | How familiar are you with home rental services? | AirBnB, HomeStay, Couchsurfing |

We estimate a series of Multinomial logit (MNL) models with varying parameter specifications, ranging from fully generic parameters (common taste parameters for the same attribute across alternatives) to fully alternative specific parameters (independent taste parameters per alternative and attribute) and dummy coded parameters, to capture potential non-linear perceptions of attributes.

As highlighted in Section 1, the joint access-mode-and-train-station choice is likely to be nested, with research being inconclusive on the overall preferred nesting structure (mode-first or station-first). Given the structure of our data (considering all ten alternatives in a single choice set), nesting of alternatives is also likely to occur. To capture potential nesting and cross nesting effects, we estimate a series of nested logit (NL), cross-nested logit (CNL) and Error component panel mixed logit (ML) models.

In addition, MNL models are also unable to capture unobserved taste heterogeneity in the sample, nor can they account for the panel effect. Two different modelling approaches are frequently used in research, which are able to mitigate these shortfalls: the panel random parameter mixed logit (ML) model and the latent class choice model (LCCM). The former extends the MNL model by allowing the taste parameters to be drawn from a distribution, acknowledging that different respondents may use different weights for the respective attributes. The latter model creates several discreet (latent) segments, each with their own parameter estimates. Both modelling approaches have their benefits and drawbacks. ML models are more parsimonious, capturing the sample heterogeneity with a relatively small number of parameters. By means of error components, the ML model is also able to capture nesting effects, as mentioned previously. LCCMs on the other hand require a larger number of parameters to be estimated, but result in a discreet number of classes, which provide a straightforward interpretation of the different population segments (Greene and Hensher 2003; Hess 2014) allowing us to distinguish segments within the population, each with its own mode preferences, time- and cost-sensitivity etc.

Another benefit of LCCMs is that the class membership function (used to determine the probability of each individual belonging to a specific segment) may include socio-demographic and attitudinal information of the respondents (Greene and Hensher 2003).

The class membership function aims to divide the population into segments that are as different from each other as possible, in order to capture the heterogeneity in the obtained SP data. By including socio-demographic information in the class membership function, more information is available on the influence of socio-demographics on class membership.

As our goal is to identify distinct user groups within the population, based on their train station access behaviour and the potential use of FLEX services, we opt for the Latent class choice model structure. Given the present nesting structure, a nested logit model is also applied within each of the segments. For capturing nesting, the NL specification is chosen over the ML for its closed-form and ease of interpretation. The resulting nesting parameter $\mu$ gives information on the level of nesting of the alternatives within the same nest. A lower value indicates the alternatives are largely independent from one another, whereas a higher value (upper bound set to 10 in models) indicates a strong nesting effect. Based on insights from the various MNL, NL and ML models estimated, a Latent class choice model with nesting structures is specified and estimated, the results of which are elaborated on in the following section.

For the class membership function of the LCCM, both socio-demographic and attitudinal information is used in the class membership. To simplify and narrow down the number of parameters, as well as to test for possible correlations between the various attitudinal statements, an exploratory factor analysis (EFA) is performed. This also allows us to test the attitudinal categories which were used when formulating the statements. The EFA is performed using the 'factor_analyzer' package for Python by Briggs (2019).

### 2.3. Data collection

The survey was distributed to participants of the Dutch Mobility Panel (MPN) (Hoogendoorn-Lanser, Schaap, and Oldekalter 2015) between February 10th and March 1st in 2020, resulting in a total of 1193 responses. The data was then processed and responses that were either (1) incomplete, (2) completed in fewer than five minutes or (3) chose the same response to all attitudinal statements, were removed from the dataset, leaving a total of 1076 responses.

The sample is largely representative of the Dutch population (Table 4). The sample displays a slight overrepresentation of older individuals, those having a higher level of education and single-person households. The difference in household income is largely due to respondents having the option not to disclose their household income (not knowing or not wishing to share that information). We believe these slight disparities to not significantly influence the model outcomes.

With respect to COVID-19, the first patient in the Netherlands was diagnosed on the 27th of February (Rijksinstituut voor Volksgezondheid en Milieu (RIVM) 2020) and the first lockdown measures were announced on March 12th (NOS 2020). We therefore believe that it is unlikely that the epidemic influenced the decision-making of the respondents.

## 3. Results

In this section, we report the survey results, outcomes of the model estimation and the interpretation of individual population segments obtained through the latent class choice model estimation. We start by summarising the descriptive statistics of the choices made by respondents in section 3.1. Then, we analyse the attitudinal statements and familiarity of the respondents with services of the sharing economy. We perform an Exploratory Factor Analysis (EFA) on the attitudinal statements, to uncover potential correlations between them, as well as to narrow down the number of attributes for the following step. This is outlined in Section 3.2. In Section 3.3, we present the results of the latent class choice model and describe the taste, attitudinal, behavioural and socio-demographic characteristics of each segment.

**Table 4.** Socio-demographics of the sample and the Dutch population (Centraal Bureau voor de Statistiek 2020).

| Variable | Level | Sample | Population |
|---|---|---|---|
| Gender | Female | 53% | 50% |
| | Male | 47% | 50% |
| Age | 18–34 | 22% | 27% |
| | 35–49 | 22% | 23% |
| | 50–64 | 30% | 26% |
| | 65+ | 26% | 24% |
| Education[a] | Low | 25% | 32% |
| | Middle | 39% | 37% |
| | High | 36% | 32% |
| Household income[b] | Below average | 21% | 26% |
| | Average | 48% | 47% |
| | Above average | 6% | 27% |
| | Did not disclose | 25% | 0% |
| Employment status | Working | 51% | 51% |
| | Not working | 49% | 49% |
| Urbanisation level | Very highly urban | 23% | 24% |
| | Highly urban | 31% | 25% |
| | Moderately urban | 17% | 17% |
| | Low urban | 21% | 17% |
| | Not urban | 8% | 17% |
| Household size | One person | 22% | 17% |
| | 2 or more | 78% | 83% |

[a]Low: no education, elementary education or incomplete secondary education; Middle: complete secondary education and vocational education; High: bachelor's or master's degree from a research university or university of applied sciences.
[b]Below average: below modal income ( < €29,500); Average: 1-2x modal income (€29,500–€73,000); Above average: Above 2x modal income ( > €73,000).

### 3.1. Descriptive statistics

The most commonly selected access modes are the bicycle and public transport (bus, tram or metro), each being selected in 36% of the cases for both the local and distant station categories. As expected, cycling dominates for accessing the local station, representing half of all choices, whereas public transport is the preferred access modes of respondents for accessing stations that are further away (beyond a comfortable cycling distance for many). All other modes (car, private and shared FLEX) are also more popular for the more distant stations. Private FLEX does not seem to be very popular as an access mode, regardless of the distance. This is not entirely surprising, given the relatively high travel cost. Shared FLEX on the other hand, seems to be reasonably attractive for accessing more distant stations, reaching a share of about 10%.

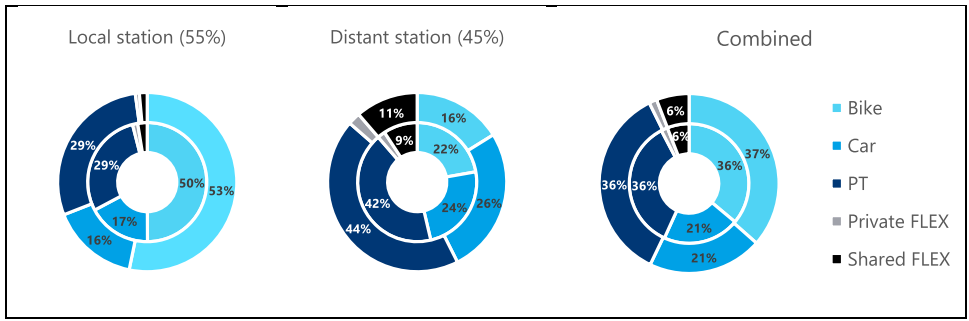

**Figure 2.** Modal split of the initially selected access modes (inner ring) and access modes selected for the actually chosen station (outer ring), for the local and distant station separately and combined.

In Figure 2, we compare the modal split for access modes that are chosen initially (inner ring), and access modes for when the corresponding station is actually chosen. The differences seem to be quite minor, with the local station seemingly being more appealing when accessed by bike, whereas the distant station was more often chosen if a motorised means of transport was selected as the access mode. Aggregating the choices between the two stations, the access mode split between the initial choices and final choices is almost identical.

## 3.2. Attitudinal statements and service familiarity

The distribution of responses and the average of each of the 16 statements, relating to the use of on-demand services, are presented in Figure 3. The first four statements capture the technology- and app-related attitudes, showing that the biggest barrier seems to be making purchases with smartphones, with the majority not willing to do so. The travel-related attributes (statements 5–8) show that people generally do not mind travelling a bit longer, provided they can use that time productively. Regarding their willingness to share (statements 9–11), respondents say they are willing to share a ride only if they get a discount, yet the proximity of strangers does not seem to be an obstacle for sharing. This could mean that sitting next to strangers is not the key reason for not pooling, but rather other aspects such as a longer and more uncertain travel (and waiting) time. For the statements on sharing economy in general (statements 12–16), people seem to be less optimistic about it for themselves, but think of it as very beneficial for society, while also seeing it as potentially leading to controversial business practices.

Similarly to what was found by Geržinič et al. (2022), the most known and often used sharing economy service in the Netherlands is food delivery, with almost half of the sample having used it at least once (as seen in Figure 4). Ride-hailing services such as Uber are familiar to most respondents, but have only ever been used by few. Most striking is that flexible public transport services, although present in several areas around the Netherlands, are unfamiliar to over half of the population. Similar results have been reported in other studies on the topic of flexible public transport (Arendsen 2019; Bronsvoort et al. 2021).

An exploratory factor analysis is performed on the 16 attitudinal statements. 'No opinion' responses are recoded to match the 'Neutral' response, to ease the performance of an EFA. This is not ideal and it is not possible to state that those responses are equivalent. Performing an EFA would thus require the removal of all respondents who at least once chose the 'No opinion' response. This results in reduction in sample size of over a third and it also inhibits the use of these factors in the class membership function of the latent class choice model. From the recoded data, we compute the KMO score to be 0.733, which indicates the sampling is middling, but still sufficient to perform an EFA (Ledesma et al. 2021).

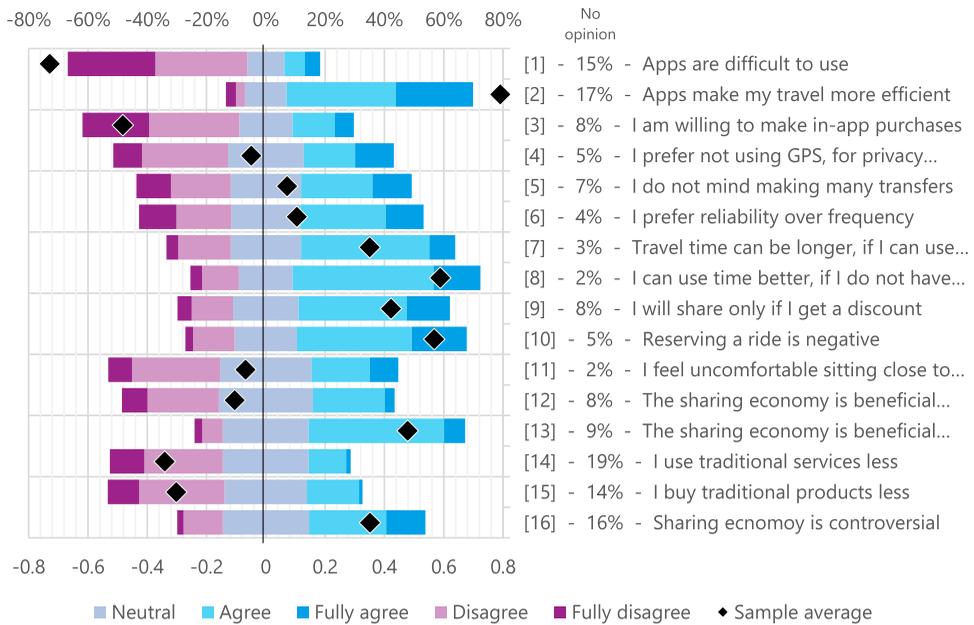

**Figure 3.** Results of the attitudinal statements.

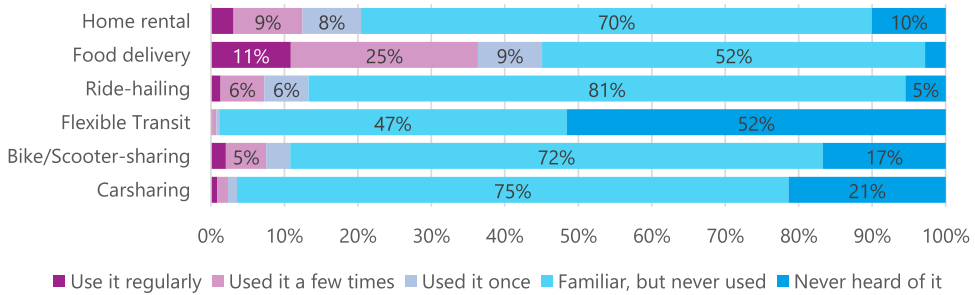

**Figure 4.** Familiarity with different sharing economy services.

Using a scree plot, we determine the optimal number of factors to be four. The resulting factor loadings onto the four corresponding factors are presented in Figure 5. The grouping of statements is largely in line with their category as indicated in Table 2. Interestingly, S5 (on making transfers) seems to be more correlated with statements on the use of travel planning apps rather than the travel related statements of S6-S8. Statement 9 (willingness to share only for a discount) and Statement 16 (controversial business practices of sharing economy) load overall weakly onto any of the four factors. The four factors can be described as 'Support for sharing economy' (F1), 'App savviness' (F2), 'Efficient travel time use' (F3) and 'Dial-a-ride scepticism' (F4).

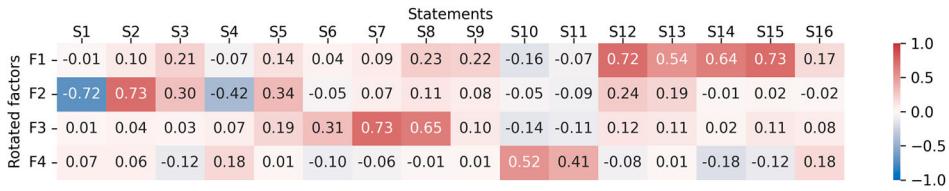

**Figure 5.** Exploratory factor analysis factor loadings for the four factors.

Based on the results of the EFA, a confirmatory factor analysis (CFA) is also performed. Factor loadings of above 0.3 (below −0.3) from the EFA are considered. The outcomes (loadings) of the CFA are used to calculate the factor values for each respondent.

### 3.3. Market segmentation

To understand how people's preferences differ, we estimate a series of latent class choice models. We choose to present a model with two sets of taste parameters, diving the population into two segments. To capture different possible nesting structures, we further divide each of the two segments into two more, where one is given a mode-based nesting structure and the other a station-based nesting structure. Both structures are presented in Figure 6. This results in a total of four segments, with two pairs sharing the same taste parameters (segments 1 and 2 vs. segments 3 and 4), and two different pairs sharing the same nesting structure (segments 1 and 3 vs. segments 2 and 4). Unlike with taste parameters, we do not restrict the nesting parameters ($\mu$) to be the same across the segments that share a nesting structure, but allow them to be estimated independently, resulting in four groups of nesting parameters, two for mode- and two for station-based nesting. This allows us to observe the different levels of correlation within the same nesting structure but among respondents with different tastes. The segmentation structure is shown in Table 5. The full set of outcomes from the model, including the model fit, taste parameters, nesting parameters and class allocation parameters are shown in Table 6. A brief overview of some estimated nested and cross-nested logit models is presented in Appendix B.

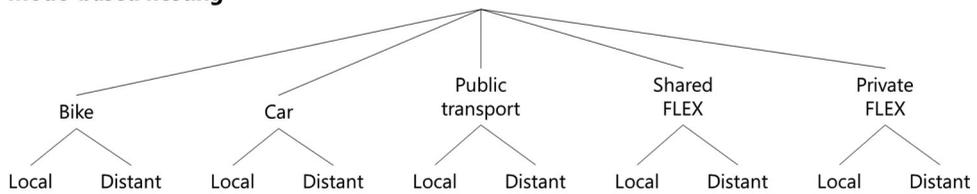

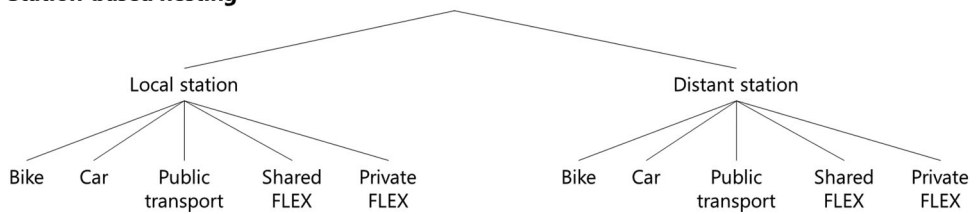

**Figure 6.** Mode-based and station-based nesting structure of choices.

**Table 5.** Segmentation structure and corresponding segment sizes.

|  | Higher WtP | Lower WtP | Σ |
|---|---|---|---|
| Mode-based nesting | *Segment 1* **21.6%** | *Segment 3* **30.2%** | **51.8%** |
| Station-based nesting | *Segment 2* **25.9%** | *Segment 4* **22.3%** | **48.2%** |
| Σ | **47.5%** | **52.5%** | **100.0%** |

The utility function used in the LCCM model is determined based on various different MNL model specifications. It is decided to use a single cost parameter for both the access leg and the train leg. In-vehicle times on the access sections of the journey are modelled as generic (common parameter for bike, car, PT and FLEX), with a second in-vehicle time parameter for the train leg. Out-of-vehicle times (OVT) are modelled using three parameters; for (1) bike and car (parking search time and walking time), (2) public transport (walking and waiting time) and (3) FLEX (waiting at home). Most of the estimated taste parameters are highly significant, with the only major exception being the waiting time for FLEX services, which is highly insignificant for all segments. It should be noted that this is the case in all the tested model specification, indicating that almost all respondents are largely indifferent to it. Similar results are also reported by Geržinič et al. (2022). This may be due to the waiting taking place at home and thus may be similar to hidden waiting time, where individuals undertake other activities at home (quick errands) while waiting. The only other insignificant parameter is the ASC for Shared FLEX for Segments 3 and 4, indicating that for them, there does not seem to be significant inherent (dis)preference for Shared FLEX over the bicycle.

Estimating the class membership parameters, an initial 4-class model with several factors and socio-demographic variables is estimated. As this results in many insignificant parameters, the characteristics which result in insignificant parameters for all classes are removed one by one, until only those remain, where a significant parameter is obtained for at least one class. Through this approach, education level, household income, urbanisation level, car ownership and train usage are removed, as well as the app savviness factor.

Each segment is presented in more detail in the following sections. For ease of interpretation, parameter trade-offs (such as the Willingness-to-Pay) are summarised and presented in Table 7. To better characterise the different segments and distinguish them from each other, their weekly travel behaviour and overall socio-demographic characteristics are presented in Figure 7 and Table 8, respectively.

### 3.3.1. Segments 1 & 2: higher WtP

Members of these segments tend to have a stronger sensitivity to travel time, particularly to in-vehicle time (IVT). They see the in-vehicle time on the access leg particularly negatively and are willing to travel more than 1.5 min longer by train to save 1 min on the access leg. Out-of-vehicle times (OVT) on the access leg are not seem that undesirable, with the

**Table 6.** Model fit, estimates of the taste, nest and class allocation parameters.

| Model fit | |
| --- | --- |
| Null LL | −14,627.17 |
| Final LL | −9,699.74 |
| Adj. Rho-square | 0.3327 |
| BIC | 19,825.31 |

| | Taste parameters | | | |
| --- | --- | --- | --- | --- |
| | Segments 1 & 2 | | Segments 3 & 4 | |
| | 47.5% | | 52.5% | |
| *Class size* | Est. | *t*-stat | Est. | *t*-stat |
| *Constants* | | | | |
| Bike | 0 (fixed) | | 0 (fixed) | |
| Car | −2.08 | −9.79*** | 1.17 | 3.43*** |
| Public Transport | −1.56 | −11.30*** | 0.85 | 3.81*** |
| Shared FLEX | −3.27 | −7.70*** | −0.06 | −0.36 |
| Private FLEX | −6.43 | −6.58*** | −0.75 | −2.44** |
| Local station | 0 (fixed) | | 0 (fixed) | |
| Distant station | −0.56 | −2.73*** | −0.44 | −2.10** |
| *Common parameters* | | | | |
| Cost | −0.28 | −7.71*** | −0.18 | −8.22*** |
| *Access leg* | | | | |
| In-vehicle time | −0.10 | −8.89*** | −0.04 | −6.84*** |
| Park & walk [bike, car] | −0.07 | −7.60*** | −0.08 | −5.37*** |
| Walk & wait [PT] | −0.04 | −1.97** | −0.02 | −2.03** |
| Wait time [FLEX] | 0.03 | 0.64 | 0.00 | 0.27 |
| *Train leg* | | | | |
| In-vehicle time | −0.06 | −10.20*** | −0.04 | −5.84*** |
| Headway | −0.04 | −7.61*** | −0.04 | −7.11*** |
| Transfer | −1.02 | −7.58*** | −0.85 | −4.77*** |

| | Nesting parameters and class allocation parameters | | | | | | | |
| --- | --- | --- | --- | --- | --- | --- | --- | --- |
| | Segment 1 | | Segment 2 | | Segment 3 | | Class 4 | |
| | 21.6% | | 25.9% | | 30.2% | | 22.3% | |
| *Class size* | Est. | *t*-stat | Est. | *t*-stat | Est. | *t*-stat | Est. | *t*-stat |
| *Nesting parameters* | | | | | | | | |
| Bike nest | 10.00 | 4.65*** | | | 1.57 | 4.32*** | | |
| Car nest | 1.45 | 3.04*** | | | 1.00 | 14.30*** | | |
| PT nest | 7.43 | 1.01 | | | 10.00 | 3.43*** | | |
| Private FLEX nest | 1.00 | 0.38 | | | 1.04 | 3.42*** | | |
| Shared FLEX nest | 2.18 | 2.62*** | | | 2.82 | 4.77*** | | |
| Local station nest | | | 2.03 | 7.46*** | | | 2.38 | 6.73*** |
| Distant station nest | | | 1.00 | 5.15*** | | | 3.11 | 5.05*** |
| *Class allocation param.* | | | | | | | | |
| Constant | 4.08 | 5.07*** | 2.94 | 4.35*** | 2.22 | 3.13*** | | |
| Age | −0.57 | −3.38*** | −0.19 | −1.55 | −0.42 | −3.13*** | | |
| BTM use | −1.12 | −5.58*** | −0.74 | −4.78*** | −1.24 | −6.86*** | Baseline | |
| Car use | −0.23 | −1.43 | −0.38 | −2.64*** | 0.64 | 3.50*** | | |
| DRT averse | −0.01 | −0.10 | 0.14 | 1.81* | 0.00 | 0.00 | | |
| SE positive | 0.39 | 2.92*** | 0.24 | 2.31** | 0.11 | 1.03 | | |
| TT use | −0.21 | −1.80* | 0.00 | −0.01 | −0.30 | −2.79*** | | |

***$p \leq 0.01$, **$p \leq 0.05$, *$p \leq 0.1$.

walking, waiting and parking search times being seen as less negative than the access IVT. Compared to the other segments, they seem to be less sensitive to other aspects of train travel, as frequency and transfers are perceived less negatively, with a transfer equalling a similar penalty to 17 min of travel by train or €3.66. Overall, members of these segments prefer the bicycle, followed by PT, the car and the two FLEX options at the end.

**Table 7.** Parameter trade-offs for different segments.

|  | Segments 1 & 2 | Segments 3 & 4 |
|---|---|---|
| *In-vehicle time* | | |
| Access IVT [€/h] | 20.62 | 13.86 |
| Train IVT [€/h] | 13.14 | 12.71 |
| Ratio access/train IVT | 1.57 | 1.09 |
| *Access segment* | | |
| PT Walk + Wait [€/h] | 9.33 | 6.88 |
| Car & Bike Walk [€/h] | 15.89 | 27.93 |
| *Train segment* | | |
| Frequency [€/h] | 8.95 | 12.81 |
| Transfer [€] | 3.66 | 4.80 |
| Transfer [min] | 16.69 | 22.64 |

Respondents in Segment 1 tend to form fairly strong nests for the bicycle, PT and to a lesser extent Shared FLEX services, with their $\mu$'s corresponding to 10, 7.4 and 2.2 respectively. The two private motorised modes (car and private FLEX) on the other hand tend to be less strongly correlated, meaning that the nesting structure is weak.

For a station-based nesting structure, the nesting effect of the local station is somewhat strong ($\mu = 2$), indicating some level of correlation between access modes. This means that a new access mode will largely result in a redistribution of travellers among the modes and attract a limited number of new users to the station. Alternatively, the modes accessing the more distant station do not seem to be correlated at all, forming independent alternatives.

### 3.3.2. Segment 1: young professionals

From the class membership function, we see that this segment is composed of younger individuals, who are average car users and not likely to use BTM. Interestingly, from Table 8 we can see that they seem to have a fairly positive about DRT and the sharing economy, yet they do not see the benefits of not having to drive themselves. Considering other socio-demographic characteristics and mobility patterns, members of this segment seem to be overall quite similar to the sample. They are, nevertheless, somewhat higher educated and tend to use the bicycle more often.

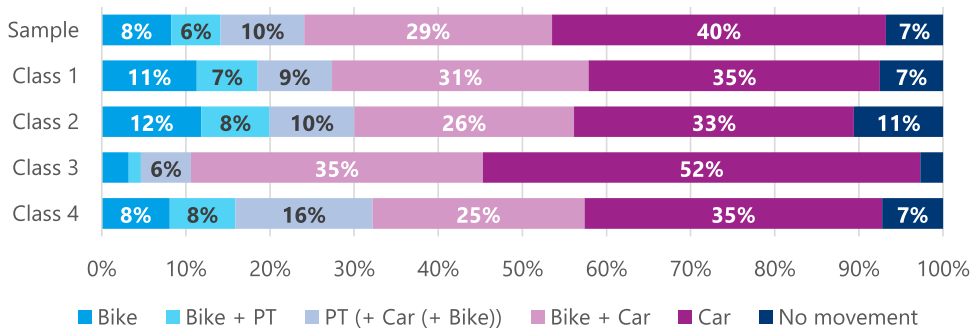

**Figure 7.** Weekly mobility patterns of the entire sample and the 4 classes (values below 5% are not labelled).

**Table 8.** Average factor scores and socio-demographic characteristics of the sample and each segment.

| | | Sample | Segment 1 | Segment 2 | Segment 3 | Segment 4 |
|---|---|---|---|---|---|---|
| Factors | Digitally challenged | – | −0.11 | 0.03 | 0.05 | −0.00 |
| | DRT averse | – | −0.13 | 0.10 | 0.12 | −0.16 |
| | Unfamiliar with SE | Normalised | −0.01 | 0.00 | 0.01 | −0.00 |
| | Positive about SE | – | 0.24 | 0.12 | −0.24 | −0.05 |
| | Effective use of TT | – | −0.03 | 0.21 | −0.34 | 0.24 |
| Gender | Female | 53% | 55% | 55% | 49% | 53% |
| | Male | 47% | 45% | 45% | 51% | 47% |
| Age | 18–34 | 22% | 29% | 19% | 22% | 16% |
| | 35–49 | 22% | 25% | 20% | 26% | 16% |
| | 50–64 | 30% | 28% | 31% | 32% | 30% |
| | 65+ | 26% | 18% | 30% | 20% | 37% |
| Education level | Low | 25% | 21% | 27% | 23% | 28% |
| | Middle | 39% | 39% | 37% | 43% | 37% |
| | High | 36% | 40% | 36% | 34% | 35% |
| Household income | Below average | 21% | 21% | 24% | 17% | 22% |
| | Average | 48% | 47% | 47% | 51% | 48% |
| | Above average | 16% | 17% | 15% | 17% | 16% |
| | Did not say | 14% | 15% | 14% | 15% | 14% |
| Employment status | Employed | 51% | 55% | 44% | 61% | 40% |
| | Student | 6% | 8% | 6% | 4% | 6% |
| | Retired | 24% | 17% | 28% | 18% | 34% |
| | other | 20% | 20% | 21% | 17% | 20% |
| Urbanisation level | Very highly urban | 23% | 22% | 25% | 17% | 30% |
| | Highly urban | 31% | 31% | 32% | 31% | 32% |
| | Moderately urban | 17% | 17% | 17% | 17% | 16% |
| | Low urban | 21% | 22% | 19% | 26% | 16% |
| | Not urban | 8% | 8% | 7% | 10% | 7% |
| Household size | 1 | 22% | 21% | 25% | 17% | 25% |
| | 2 | 36% | 31% | 37% | 34% | 41% |
| | 3+ | 42% | 48% | 38% | 49% | 34% |
| Household car ownership | Average | 1.17 | 1.17 | 1.04 | 1.38 | 1.06 |
| | 0 | 15% | 17% | 21% | 5% | 19% |
| | 1 | 56% | 53% | 56% | 57% | 59% |
| | 2+ | 29% | 30% | 23% | 38% | 22% |

### 3.3.3. Segment 2: middle-aged neutrals

Class allocation parameters for Segment 2 indicate that their members tend to be older, infrequent car users and more frequent BTM users. Similar to Segment 1, they are positive about the sharing economy, but in contrast, they do see the benefits of not having to drive themselves, yet are more DRT averse. Other factors indicate they are more digitally challenged and less experienced with using services of the sharing economy. They tend to live in more urban areas and are on average lower educated and the least affluent of the segments. They also have the lowest car ownership of any segment, with only 1.04 vehicles per household and 21% living in households without a car at all. Logically, they are also the least frequent car users and thus use public transport or cycle more often. They are also the most likely to not travel at all regularly on a weekly basis.

### 3.3.4. Segments 3 & 4: lower WtP

Compared to the first two segments, members of these tend to be slightly more cost sensitive, especially when it comes to access IVT. They do not perceive the access and train-leg IVT very differently, meaning that they would prefer to minimise their overall travel time and do not have a particular preference for one leg or the other. They are however very strongly averse to the parking search time for bike and car, seeing it twice as negative as access IVT. They are also less tolerant of transfers than the other segments, being willing to travel 5 min or paying over €1 more compared to the other segments. Their overall access mode preferences lie with the car and PT. Cycling and Shared FLEX are seen as roughly equal, whereas Private FLEX is again least preferred.

Similar as with Segment 1, the nesting of PT and Shared FLEX alternatives in Segment 3 is quite strong, indicating that the users of these modes would likely keep using them, even if a new station opened. Car and Private FLEX are also, like in Segment 1, highly uncorrelated, meaning that multiple alternatives (train stations) are independent of one another when considering these modes. The key difference however, is for the cycling alternatives, which do not seem to be very strongly correlated in Segment 3 ($\mu = 1.6$), as opposed to the very strong nesting structure in Segment 1 ($\mu = 10$).

In Segment 4, nesting for both station alternatives tends to be reasonably strong, with both having a $\mu$ of over 2. The correlation of access modes to the local station is similar as in Segment 2, whereas much stronger nesting for the distant station can be observed in Segment 4.

### 3.3.5. Segment 3: exurban car drivers

Members of Segment 3 tend to be very frequent car users and thus infrequent BTM users. They are the most DRT averse, digitally challenged, least positive about the sharing economy and also see the least benefits in not having to drive themselves. They are predominantly young adults, with the largest share of individuals obtaining a middle education and an average to above average income per household. They are also the most likely to live in large household (three or more people). Being the least likely segment to live in an urban area (rather living in suburban and rural areas), it is logical that they are the segment with the highest car ownership (almost 1.4) and only 5% do not have a car at all. 62% also drive their car daily, compared to 40% in the sample average. This last also corresponds to their very low use of other travel modes.

### 3.3.6. Segment 4: urban PT enthusiasts

Conversely to the previous, members of Segment 4 are frequent BTM users and less frequent car users. They are also more SE averse, but tend to be positive towards DRT and the use of travel time for other activities. They are overall the oldest of the four segments, with an average income and a below average level of education. Corresponding to their lower car use, their car ownership is below average and almost 19% do not own a car at all. In addition to being the most frequent BTM users (20% on a weekly basis, compared to 11% in the sample), they are also above average train users, making them the overall strongest PT users.

## 4. Model application: scenario analysis of market potential

In the following we evaluate how the introduction of FLEX and the variation if its service level impacts modal split and travel behaviour. Firstly, we look at different FLEX introduction scenarios and how the market shares between modes shift due to this introduction. Secondly, we vary several attributes of the trip, including (1) the distance of the more distant station, (2) the average speed of FLEX and (3) the number of transfers saved by travelling via the distant station. We evaluate the impact of this on the individual class level and at an aggregate level. As a baseline, we take a typical medium-distance trip with two possible stations to access and four access modes for each. The attribute levels are presented in Appendix C in Figure 13. The assumed average travel speeds for calculating the travel times of the access modes are 15 km/h for the bicycle, 24 km/h for the car and 18 km/h for public transport and 20 km/h for FLEX.

We also carry out a sensitivity analysis, the full results of which are reported in Table 10 in Appendix D. Overall, we observe that demand (market share) is largely inelastic. Individuals seem to be most sensitive to the access leg in-vehicle time. Interestingly, the ticket price of a longer public transport access trip is quite elastic, resulting in a shift in demand that is greater than 0.1 for many modes. Demand for existing transport modes (bike, car, PT) tends to be more sensitive to time, whereas the demand for FLEX seems to be more sensitive to price. It should be noted that the biggest changes in demand occur for alternatives that have an initially small market share, as a slight increase in demand means a big proportional change in market share.

### 4.1. Introducing an on-demand service

We apply the outcomes of the choice model to examine how the existing modal split is affected in four introduction scenarios of FLEX. Two scenarios model a 'Competition' style entry of FLEX, acting as a direct competitor to existing services. The other two scenarios consider a 'Substitution' setting in which FLEX replaces PT services in the study area. As our interest is also the interaction of access mode and station choice, we also consider if the more distant station is new alternative and its opening coincides with the introduction of FLEX, or if it is already present when FLEX is launched. The impacts of the scenarios on the modal shift are presented in Figure 8.

In all four scenarios, we can observe that FLEX obtains a fairly small market share; approximately 7% in the *Competition* scenario and 12% in the *Substitution* scenario. In the former, the split between the two stations is about equal, whereas in the latter, almost two thirds of the FLEX trips are made accessing the local station, despite the distant station having an overall higher market share. Interestingly, in the case of a new station opening, most FLEX users are former PT users, whereas if both stations are already present, they are primarily former car drivers (specifically distant station car drivers). In either case, cyclists do not really shift in large numbers to using FLEX. Considering the impact of FLEX on overall station market share, it is almost insignificant, only marginally adding to the attractiveness of the local station (change in market share is less than one percentage point).

Turning next to the impact of PT substitution, the first and clear impact is an increased share of both car (10 percentage points) and, to a lesser extent, bike use (six percentage points), with FLEX providing an alternative for only a small number of former PT users. Despite it often being touted as a PT replacement, our results seem to suggest that this is not as straightforward. When only one station is present at the start, most FLEX passengers are former PT users (approximately 60%). In the case of both stations already being present, most FLEX users (again) switch from their car however (45%), with limited correlation between the local and distant stations. Looking at where former PT users shift to, it is mainly to the car (especially to access the distant train station), with cycling to the local station being an attractive option primarily for travellers already using the local station.

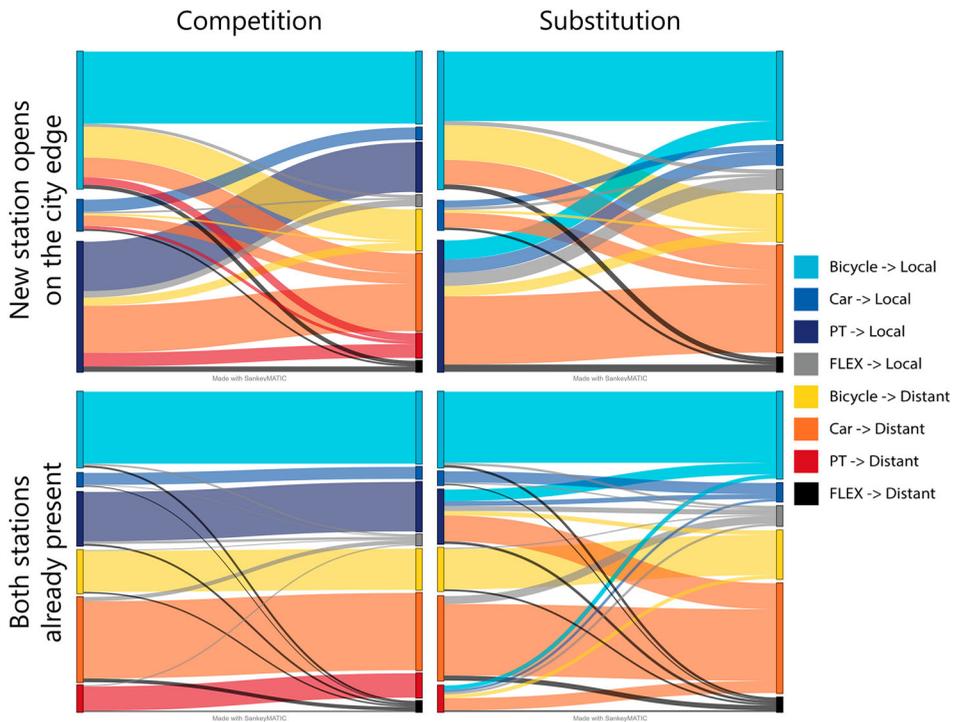

**Figure 8.** The impact of the Introduction and Substitution scenarios on modal split for train station access.

Mode and station market shares used in this example are heavily dependent on the selected attribute levels. Nevertheless, we can see that FLEX is not a highly competitive alternative, capturing only a small share of the market. If a new station opens some distance away from the existing one, where cycling becomes too strenuous for most, FLEX can provide a viable alternative, although still representing a small share, compared to the car, which dominates as the access mode to the new station. The impact of distance on the attractiveness of FLEX, along with varying other operational characteristics, is investigated in the following section.

## 4.2. Level-of-service variation

Figures 9 and 10 show the scale of changes in the market share when varying FLEX travel speed, number of transfers saved and access distance. The latter requires some further clarification. We fix the access distance to the local station at 3 km from home and then vary the additional travel distance to the distant station. The D distance in both Figures 9 and 10 thus indicates the extra distance (varied from 0 to 7 km farther), meaning that the total access distance is varied between 3 and 10 km. The trip characteristics are identical to what is shown in Figure 13, where the distant station is 8 km away from home (5 km farther than the local station).

In both figures, we see that FLEX is less attractive for shorter distances, becoming an increasingly attractive alternative with the distance becoming too long for most to cycle, plateauing somewhere around 7–8 km away from home (4–5 km farther than the local station). Segment 3 (Exurban car drivers) seems to be most likely adopter of FLEX services, with a market share of around 15%, whereas the remaining three segments at around 5% market share. In terms of sensitivity to the distance, Segment 1 (Young professionals) seem to be the most affected by it, switching almost entirely to the local station if saves them 5 km or more of travelling. The least sensitive on the other hand are members of Segment 4 (Urban PT enthusiasts). Looking at the combined market share of the distant station, an interesting observation can be made at around 4–5 km of extra travel distance. That seems to be the tipping point for the overall population on which station to choose. A likely reason is that at that point, cycling becomes too unattractive to access the local station and bike nesting individuals (Segments 1 and 3) make the switch to the local station at roughly that distance.

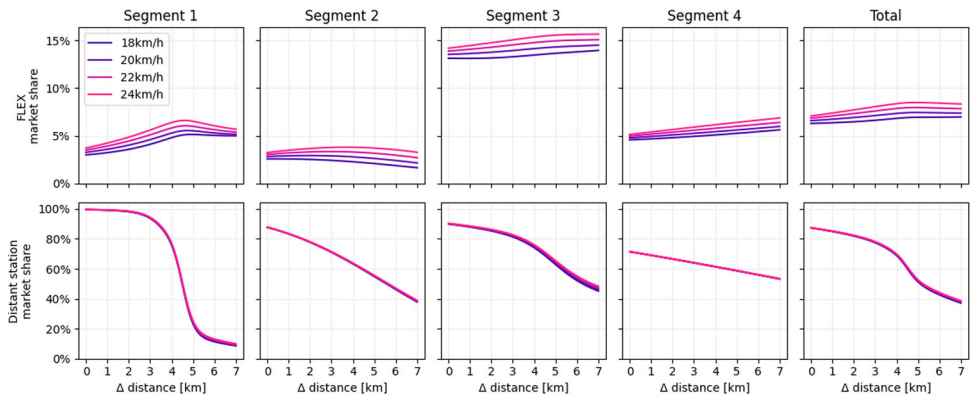

**Figure 9.** Market share for FLEX and Distant station when varying the average travel speed of FLEX and the distance between the two stations.

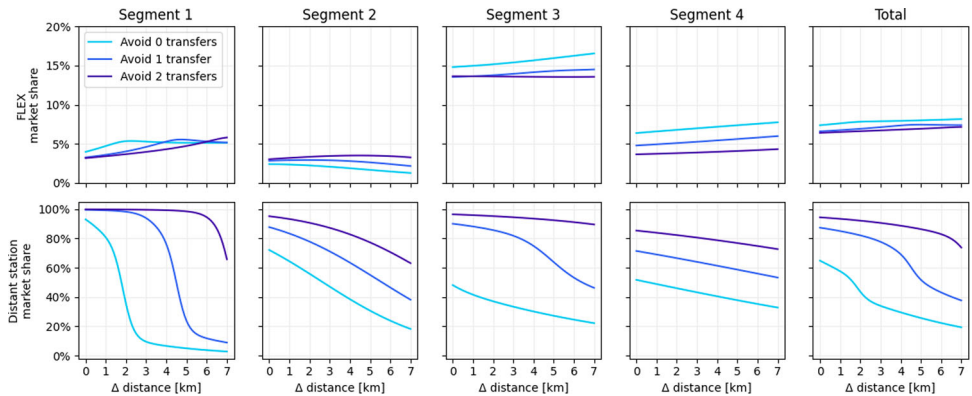

**Figure 10.** Market share for FLEX and Distant station when varying the number of transfers saved by travelling via the distant station and the distance between the two stations.

Considering the varying FLEX speed, a higher speed does increase overall FLEX market share, although not more than one percentage point. Because the speed increases for accessing both stations, the impact on station choice is very limited, with only a marginal increase for the distant station at higher speeds. We assumed an average speed of 20 km/h, slower than car as FLEX has to potentially make additional stops and detours to pick up other passengers, yet we still consider it faster than PT, as it does not stop that often. The average FLEX speed can be influenced by allowing the vehicles to use PT lanes, giving them priority at traffic lights and by determining the longest allowed detours for picking up additional travellers. Given the relatively minor changes to market share, it may be more beneficial to pick up additional passengers rather than use more vehicles to guarantee a quicker trip, for a marginal improvement in attractiveness.

The second analysis focuses on saving transfers on the train leg of the trip. Transfer provide a significant barrier in train travel for many passengers. Our results also support this notion, with a transfer being perceived equally as 15–25 min of travel time or €3–5 of trip costs. FLEX market share seems to overall decrease when the local station has additional transfers, which is somewhat logical, given that the local station was more attractive for FLEX users. The change in market share is again quite limited, although an interesting pattern can be observed for Segment 1. The attractiveness of FLEX peaks at a greater distance if more transfers can be saved. The market share also increases with distance, which likely due to the bicycle becomes a less viable alternative. Turning to the station market shares, we see the impacts are quite significant. As in Figure 9, Segment 1 is highly sensitive to distance, with the number of transfers saved only influencing at what distance they would shift. The results indicate that for saving a transfer, they are willing to travel approximately 3 km farther. Big differences (of at least 20% points per transfer saved) can also be observed for other segments, although their sensitivity to distances is less pronounced. A particularly high sensitivity can be observed in Segment 3, where saving two transfers makes a difference of almost everyone or no one from that segment using the distant station.

## 5. Conclusion

In this paper, we explore the potential of using on-demand mobility services (FLEX) for home-end first/last mile train station access in the Netherlands. Improving station access is an essential aspect in increasing train use and is as important as improving the train service itself. We analyse the joint choice of access mode and train station, by applying a sequential stated preference survey design, disseminating it through the Dutch Mobility Panel (Hoogendoorn-Lanser, Schaap, and Oldekalter 2015). We estimate several choice models in order to examine the prominence of access station versus access mode choice, user heterogeneity and market segments. Here, we first present and discuss the main findings (Section 5.1). This followed by policy implications of introducing FLEX services in Section 5.2, before finalising with the limitations of this research and outlining the future outlook (Section 5.3).

### 5.1. Discussion and key findings

Model estimates show that respondents prefer the existing access modes, such as the bicycle, car and public transport, over on-demand services. This is in line with other studies analysing the potential of on-demand mobility (Frei, Hyland, and Mahmassani 2017; Geržinič et al. 2022; Liu et al. 2018), possibly due to the unfamiliarity of respondents with novel services. A generic IVT parameter for the access leg shows that respondents perceive it more negatively than the main leg travel time (Arentze and Molin 2013; Bovy and Hoogendoorn-Lanser 2005; La Paix Puello and Geurs 2014), although we show that the scale of the difference in perception varies between users. A somewhat unexpected finding is the perception of waiting time for on-demand service, which seems to be insignificant. Arguably, this is due to a combination of its description in the survey – as waiting time is presented as waiting at home – and the small attribute levels used, ranging between one and nine minutes. A similar result was found in our previous study on on-demand services for urban travel (Geržinič et al. 2022). The potential to have more time to get ready or to run a quick errand before leaving is presumably the reason for such an estimate. We suspect that a negative perception would be observed if longer waiting times would have been used or if the waiting would have to occur at the pick-up location on the street.

By means of a latent class model, we uncover and characterise four distinct user groups, based on their taste heterogeneity (time–cost trade-offs), nesting structure, socio-demographic characteristics, mobility patterns and attitudinal statements. Of the uncovered segments, the lower WtP segments, in particular the mode-based-nesting segment (*Exurban car drivers*) seems to be the most likely to adopt FLEX services for station access. Although they prefer the car and PT for station access, shared FLEX is not seen as significantly inferior to the bicycle for example. Their aversion to transfers on the train leg, higher sensitivity to train frequency and an almost identical weighing of IVT on the access and main leg also make them more likely to opt for a more distant station if it provides a superior service. The two segments with a higher WtP seem to be less likely FLEX adopters. Although they use their cars less intensely and have a higher value of time, they have a strong preference for cycling and a strong dispreference for FLEX. In addition, their stronger penalty for the access leg IVT means they are more likely to access a nearby station if possible and are willing to tolerate more travel time and transfers while travelling by train. This means they are more likely to opt for a local station, which, in combination with their high cycling preference, means that they are much more likely to cycle overall.

With respect to the two possible nesting structures, we show that both are almost equally prevalent amongst respondents, with the segments modelling mode-based nesting representing roughly 52% of the population, whereas the station-based nesting segments account for 48%. This is largely in line with previous literature, which reports mixed results (Bovy and Hoogendoorn-Lanser 2005; Chakour and Eluru 2014; Debrezion, Pels, and Rietveld 2009; Fan, Miller, and Badoe 1993). As the difference between the two segments is fairly minor in our results, a small change in context is likely to tilt the model performance and favour one or the other nesting structure.

With the two-level segmentation structure (two pairs of segments sharing the same taste parameters and then two more based on mode nesting), it is interesting to see similarities in attitudes and socio-demographics in segments with similar WtP and with the same nesting structures. When grouping segments based on their taste parameters, individuals with a higher WtP tend to have a more positive view of the sharing economy. When analysing similar nesting structures, respondents who tend to nest alternatives based on the train station tend to be older, have a slightly lower level of education, more urban dwelling and living in smaller households. They also see the benefit of not having to drive themselves and being able to use that time productively. Interestingly, there are some similarities that can also be observed between two diagonal segments that have both different taste parameters and nesting structures; Segments 1 (high-WtP, mode-nesting) and 4 (low-WtP, station-nesting) as opposed to Segments 2 and 3. The former two nests tend to have more experience with services of the sharing economy, are more tech savvy and have a more positive view towards DRT services.

The four clusters show similarities to other studies looking into market segmentation with respect to new mobility solutions (Alonso-González et al. 2020a; Alonso-González et al. 2020b; Geržinič et al. 2022; Winter et al. 2020). Most of these studies report at least one group that is largely ready to adopt mobility innovations and is currently fairly multimodal in their travel behaviour. In our study, this is somewhat split between Segments 1 and 4, where the former is more ready to adopt mobility innovations, whereas the latter is the most multimodal of any. These two segments also relate strongly to two further often uncovered groups, with one being a technologically-savvy car driving segment also shows potential for innovation adoption, but they tend to be time-sensitive (comparable to Segment 1). The other typical segment is a public transport supporting cluster, which tends to be more cost-sensitive and largely willing to adopt innovation, but are somewhat limited due to their cost-sensitivity (largely in line with Segment 4). Finally, most studies also find a segment in the population that is more negative/reluctant towards the adoption of innovations and also prefers to drive a car (very similar to Segment 3). It is interesting to point out however, that despite these characteristics of Segment 3, they seem to be a strong contender for FLEX adoption, based on our findings. The aforementioned studies are all based on separate data collection efforts, samples and models estimated, hence the uncovered parallels to these studies further support the findings of this research.

## 5.2. *Policy implications*

Applying the model estimates, we show that introducing an on-demand service will not have a significant impact on any existing mode, with most users coming from the car if FLEX is added as an additional service. Although not directly resulting from our study, we speculate that some travellers would likely not travel at all if public transport was entirely substituted by on-demand services.

If implemented, on-demand services would capture a fairly niche market, attracting users away from PT and car as an access mode to train stations. To limit the modal shift from public transport as much as possible, the planning of fixed (traditional) and flexible (on-demand) public transport should be integrated. Pinto et al. (2020) show that both the users and the operators could benefit from jointly planning (re-designing) a public transport network made up of fixed lines and flexible services. The greatest benefits of replacing fixed lines are likely to stem from current low-demand areas, where PT is operated at low-frequencies and is therefore less attractive to users. Notwithstanding, the results of Pinto et al. (2020) and also those of Narayan et al. (2020) suggest that ridership of fixed PT would nevertheless decline.

With respect to operational characteristics, FLEX services should aim at bundling multiple travellers into a single vehicle, reducing the overall vehicle miles travelled. This can however lead to more stops and detours, increasing the overall trip time and reducing the average speed. To counteract that, services can be given priorities reserved for public transport, such as the use of dedicated lanes and priority at traffic lights. Our results show that the travel speed does not have a significant impact on the attractiveness of the service and thus on the market share. Orienting the service to pick up a larger number of passengers, at the expense of a few minutes of travel time might therefore be reasonable. Designated pick-up and drop-off locations, with potentially similar amenities as bus stops, may also help reduce the scale of detours necessary to pick-up passengers and thus decrease travel time, but would result in travellers having to walk a certain distance, reducing the attractiveness of the service. Given the limited sensitivity to waiting at home, pick-up and drop-off locations may be best avoided, with the higher attractiveness of waiting at home possibly compensating for the slightly longer travel time.

In terms of joint access mode and station access, we show that on-demand services do not have a significant impact on the share of one particular station, with FLEX being an equally attractive service to both more local and more distantly located stations. It is important to note however, that when serving more remote stations, FLEX tends to compete primarily with the car as the access mode, whereas when serving a local station, the key competitor is public transport.

### 5.3. Limitations and future research

Our research utilises a stated preference approach, which allows us to investigate the attitudes and perceptions of travellers towards services that are not yet widespread and/or commonly known by the local population. However, this does bring with it the limitations associated with SP studies, namely hypothetical bias and a potentially high willingness-to-pay displayed by respondents (Loomis 2011; Murphy et al. 2005). All respondents were also presented with two train stations to choose between, however some may not have any choice at all in reality, whereas others may have even more options to choose among, making the survey less realistic for some.

Future research can also test for the transferability of our market segmentation results to other contexts, particularly the size and composition of the segments, which we expect to differ depending on the trip purpose and geographical area. Our results are based on the attitudes of the Dutch population and Dutch context, meaning there is a high prevalence for cycling to the station and often limited car parking availability.

To understand how on-demand service can help in attracting more train travellers, an alternative to the main trip leg (train) should also be studied. It is likely that many participants would not have travelled by train if given the option, yet they were forced to choose an access mode and train station. This may have skewed the results towards a particular mode or attribute. However, it does show the preferences of the entire population, providing us with knowledge on attributes that require attention to attract all types of travellers.

Our study also did not take the activity side of the trip into account. While there is likely limited impact of the activity-end mode choice on the home-end mode choice, we cannot be certain and studying the complete trip would allow one to state with more certainty whether or not this is the case. The activity-end is also interesting to study in and of itself. As travellers rarely have their own means of mobility available on the activity side of the trip, shared mobility services may prove highly attractive and could potentially increase the share of train users.

Finally, a highly relevant characteristic of station access is reliability, both of the travel time and parking search time. If train services are not very frequent, this may be a key deciding factor for many travellers, choosing an alternative that is reliable and gives them the best chance of making their connection. Including this variability was beyond the scope of our research, but could provide invaluable insight into future service design.

## Acknowledgements

The authors thank The Netherlands Institute for Transport Policy Analysis (KiM) for facilitating the survey data collection and the Dutch Mobility Panel (MPN) participants for their time and effort in responding to the survey. The authors confirm contribution to the paper as follows: Study conception and design: Geržinič, Cats, van Oort; Survey design and data gathering: Geržinič, Hoogendoorn-Lanser, van Oort, Cats; Analysis and interpretation of results: Geržinič, van Oort, Cats; Draft manuscript preparation: Geržinič, van Oort, Cats; Supervision and reviewing: Cats, van Oort, Hoogendoorn; Funding acquisition: Cats. All authors reviewed the results and approved the final version of the manuscript.

## Disclosure statement

## Funding

This research was supported by the CriticalMaaS project [grant number 804469], which is financed by the European Research Council and Amsterdam Institute for Advanced Metropolitan Solutions.

## ORCID

*Nejc Geržinič*  http://orcid.org/0000-0001-7533-0109
*Oded Cats*  http://orcid.org/0000-0002-4506-0459
*Niels van Oort*  http://orcid.org/0000-0002-4519-2013
*Sascha Hoogendoorn-Lanser*  http://orcid.org/0000-0002-1829-2481
*Serge Hoogendoorn*  http://orcid.org/0000-0002-1579-1939

# Appendices

## Appendix A

### Survey

In the survey, respondents are asked to make 3 successive choices for making one multimodal trip. Choices 1 and 2 required respondents to choose an access mode to two different train stations at different distances from their trip origin. An example choice set is shown in Figure 11. The 'Car' alternative was only presented to the respondents who indicated they have a drivers licence and access to a car.

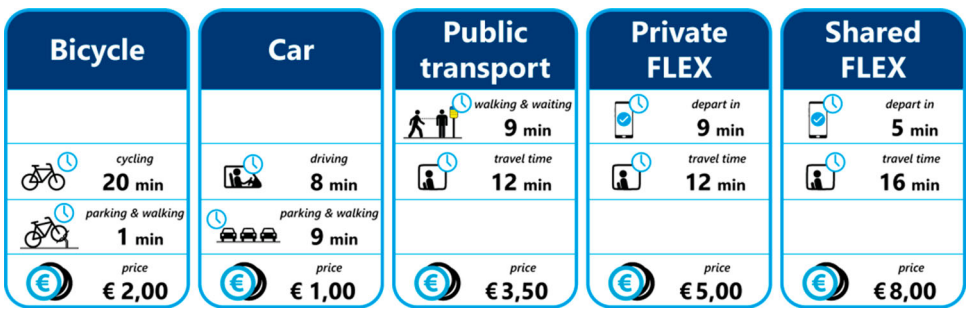

**Figure 11.** Example choice set of Choice 1 (and Choice 2).

Based on the access modes chosen in Choices 1 and 2, a choice set containing those access modes and the train service characteristics are presented to the respondents, as highlighted in Figure 12. In this particular case, the respondent has chosen to access Station A (Choice 1) by bicycle and Station B (Choice 2) by Public transport.

## Appendix B

### Cross-nested logit models

To better understand and capture the different possible nesting structures, several different cross-nested logit (CNL) models are also estimated. With multiple nesting structures possible in our dataset, CNL models are able show (probabilistically), how the alternatives are allocated to different nest structures. Compared to the MNL and NL models, CNL models are statistically superior (see Table 9). However, another possible way of accounting for this cross-nesting structure is by means of a latent class choice model (LCCM), with common taste parameters and two different nesting structures. The benefit of using LCCMs as opposed to CNL models is that we can include socio-demographic characteristics in the class allocation function, to better understand not only how the alternatives are allocated to different nesting structures, but also what are the characteristics of the individuals who form certain nesting structures. Estimating a 2-class LCCM with common taste parameters are results in a statistically superior model fit to the CNL model (see Table 9).

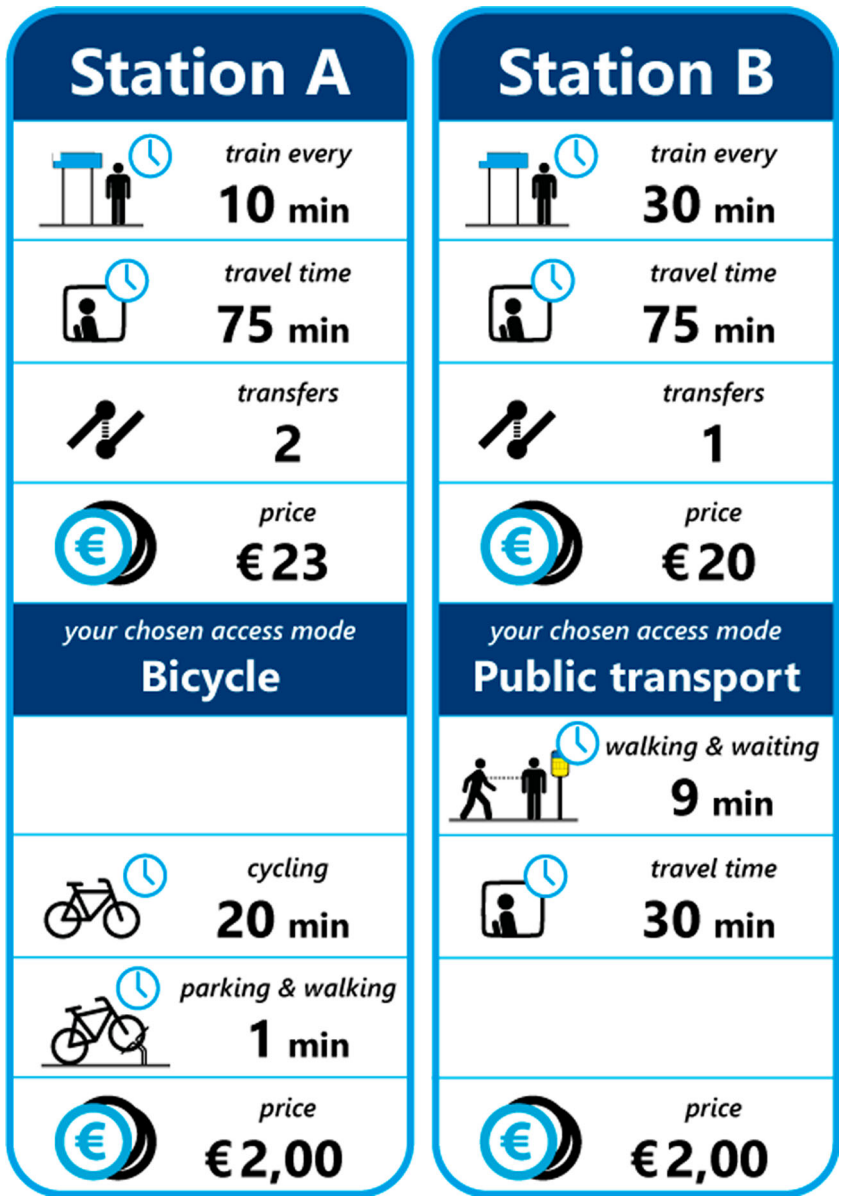

**Figure 12.** Example choice set of Choice 3.

**Table 9.** Model fits of non-nested, nested, cross-nested and latent-class nested logit models.

|  | MNL model | NL model (station-based nesting) | NL model (mode-based nesting) | CNL model | 2-class LCCM with nesting |
|---|---|---|---|---|---|
| Final LL | −11,238 | −11,134 | −11,114 | −10,853 | −10,768 |
| Adj. rho-square | 0.2308 | 0.2388 | 0.2389 | 0.2556 | 0.2623 |
| BIC | 22,590.15 | 22,408.96 | 22,394.69 | 22,014.72 | 21,689.51 |

# Appendix C

Example options for the trip.

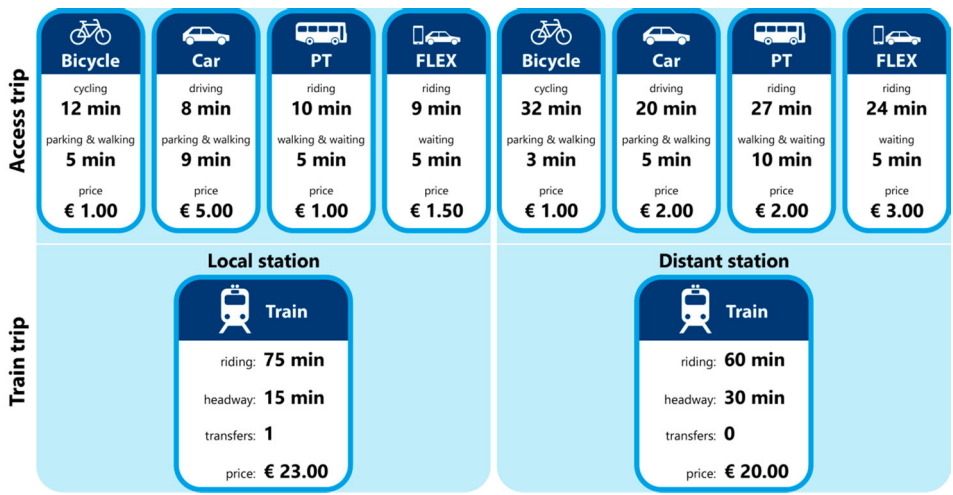

**Figure 13.** Example trip options (left): Via local train station; right via IC train station.

# Appendix D

## *Sensitivity analysis*

**Table 10.** Results of the sensitivity analysis.

| | | Local station | | | | | Distant station | | | | |
|---|---|---|---|---|---|---|---|---|---|---|---|
| | | Bike | Car | PT | FLEX shared | FLEX private | Bike | Car | PT | FLEX shared | FLEX private |
| *Initial market shares* | | 0.28 | 0.04 | 0.40 | 0.03 | 0.01 | 0.01 | 0.02 | 0.18 | 0.03 | 0.00 |
| Bike to Local station | Access time | −0.08 | 0.11* | 0.03 | 0.03 | 0.02 | 0.01 | 0 | 0.01 | 0 | 0 |
| | Parking cost | −0.02 | 0.02 | 0.01 | 0.01 | 0 | 0 | 0 | 0 | 0 | 0 |
| | Parking search time | −0.02 | 0.02 | 0.01 | 0.01 | 0 | 0 | 0 | 0 | 0 | 0 |
| Car to Local station | Access time | 0.01 | −0.08 | 0 | 0.01 | 0 | 0 | 0 | 0 | 0 | 0 |
| | Parking and trip cost | 0 | −0.05 | 0 | 0 | 0 | 0 | 0 | 0 | 0 | 0 |
| | Parking search time | 0 | −0.06 | 0 | 0 | 0.01 | 0 | 0 | 0 | 0 | 0 |
| PT to Local station | Access time | 0.03 | 0.03 | −0.05 | 0.05 | 0.01 | 0.01 | 0.01 | 0.05 | 0.02 | 0.01 |
| | Ticket price | 0.02 | 0.02 | −0.04 | 0.03 | 0.02 | 0.01 | 0.01 | 0.05 | 0.01 | 0.01 |
| | Walking & waiting time | 0.01 | 0.01 | −0.01 | 0.01 | 0 | 0 | 0 | 0.02 | 0 | 0 |
| Shared FLEX to Local station | Access time | 0 | 0 | 0 | −0.1* | 0.01 | 0 | 0 | 0 | 0.01 | 0.01 |
| | Ride price | 0 | 0 | 0 | −0.11* | 0.03 | 0 | 0 | 0 | 0.01 | 0.02 |
| | Waiting time | 0 | 0 | 0 | 0.02 | 0 | 0 | 0 | 0 | 0 | 0 |
| Private FLEX to Local station | Access time | 0 | 0 | 0 | 0 | −0.08 | 0 | 0 | 0 | 0 | 0.02 |
| | Ride price | 0 | 0 | 0 | 0.01 | −0.16* | 0 | 0 | 0 | 0 | 0.03 |
| | Waiting time | 0 | 0 | 0 | 0 | 0.01 | 0 | 0 | 0 | 0 | 0 |
| Bike to Distant station | Access time | 0 | 0 | 0 | 0 | 0 | −0.33* | 0.02 | 0.01 | 0.02 | 0.01 |
| | Parking cost | 0 | 0 | 0 | 0 | 0 | −0.04 | 0 | 0 | 0 | 0 |
| | Parking search time | 0 | 0 | 0 | 0 | 0 | −0.09 | 0 | 0 | 0 | 0.01 |
| Car to Distant station | Access time | 0 | 0 | 0 | 0 | 0 | 0.02 | −0.13* | 0.01 | 0.01 | 0.01 |
| | Parking and trip cost | 0 | 0 | 0 | 0 | 0 | 0.02 | −0.16* | 0.01 | 0.01 | 0.01 |
| | Parking search time | 0 | 0 | 0 | 0 | 0 | 0.02 | −0.14* | 0.01 | 0.01 | 0.01 |
| PT to Distant station | Access time | 0.01 | 0.02 | 0.03 | 0.01 | 0.04 | 0.12* | 0.13* | −0.12* | 0.05 | 0.13* |
| | Ticket price | 0 | 0.01 | 0.01 | 0.01 | 0.01 | 0.07 | 0.07 | −0.05 | 0.05 | 0.05 |
| | Walking & waiting time | 0 | 0 | 0.01 | 0 | 0.01 | 0.02 | 0.01 | −0.02 | 0.01 | 0.02 |
| Shared FLEX to Distant station | Access time | 0 | 0 | 0 | 0.01 | 0.01 | 0.03 | 0.03 | 0.01 | −0.1* | 0.02 |
| | Ride price | 0 | 0 | 0 | 0.01 | 0.01 | 0.03 | 0.02 | 0.01 | −0.11* | 0.03 |

**Table 10.** Continued.

| | | Local station | | | | | Distant station | | | | |
|---|---|---|---|---|---|---|---|---|---|---|---|
| | | Bike | Car | PT | FLEX shared | FLEX private | Bike | Car | PT | FLEX shared | FLEX private |
| Private FLEX to Distant station | Waiting time | 0 | 0 | 0 | 0 | 0 | 0 | 0 | 0 | 0 | 0 |
| | Access time | 0 | 0 | 0 | 0 | 0 | 0 | 0 | 0 | 0 | −0.15* |
| | Ride price | 0 | 0 | 0 | 0 | 0.01 | 0 | 0 | 0 | 0 | −0.38* |
| Travel via Local station | Waiting time | 0 | 0 | 0 | 0 | 0 | 0 | 0 | 0 | 0 | 0 |
| | Train travel time | 0 | −0.03 | −0.02 | −0.06 | −0.09 | 0.06 | 0.05 | 0.05 | 0.06 | 0.1 |
| | Train ticket price | 0 | −0.04 | −0.03 | −0.08 | −0.13* | 0.07 | 0.07 | 0.07 | 0.06 | 0.13* |
| | Operating frequency | 0 | −0.01 | 0 | −0.01 | −0.03 | 0.01 | 0.01 | 0.01 | 0.01 | 0.02 |
| | Transfers | 0 | −0.02 | −0.01 | −0.04 | −0.05 | 0.03 | 0.04 | 0.03 | 0.03 | 0.05 |
| Travel via Distant station | Train travel time | 0.01 | 0.02 | 0.02 | 0.03 | 0.06 | −0.06 | −0.04 | −0.04 | −0.06 | −0.1* |
| | Train ticket price | 0.01 | 0.03 | 0.02 | 0.04 | 0.07 | −0.08 | −0.05 | −0.06 | −0.07 | −0.13* |
| | Operating frequency | 0 | 0.01 | 0.01 | 0.01 | 0.02 | −0.01 | −0.01 | −0.01 | −0.02 | −0.03 |
| | Transfers | 0 | 0.01 | 0 | 0 | 0.01 | −0.01 | 0 | −0.01 | −0.01 | −0.02 |

Grey shade cells indicate differences greater than 0.1.